\documentclass[preprint,eqsecnum,preprintnumbers,nofootinbib,byrevtex,prd,aps,showpacs,showkeys,groupedaddress,floatfix]{revtex4}
\usepackage{bm}
\usepackage{graphics}
\usepackage{graphicx}
\usepackage{epsfig}
\usepackage{amssymb}
\usepackage{amsmath}
\begin{document}

%\large
\title{PION WAVE FUNCTIONS FROM HOLOGRAPHIC QCD AND THE ROLE OF INFRARED  RENORMALONS IN PHOTON-PHOTON COLLISIONS}
\author{A.~I.~Ahmadov$^{1,2}$~\footnote{E-mail: ahmadovazar@yahoo.com}}
\author{C.~Aydin$^{1}$~\footnote{E-mail: coskun@ktu.edu.tr}}
\author{F.~Keskin$^{1}$~\footnote{E-mail: feridunkeskin@ktu.edu.tr}}

\affiliation {$^{1}$ Department of Physics, Karadeniz Technical
University, 61080, Trabzon, Turkey\\
$^{2}$ Department of Theoretical Physics, Baku State University, Z.
Khalilov st. 23, AZ-1148, Baku, Azerbaijan}

\date{\today}
\begin{abstract}
In this article, we calculate the contribution of the higher-twist
Feynman diagrams to the large-$p_T$  inclusive single pion
production cross section in photon-photon collisions  in case of
the running coupling and frozen coupling approaches within
holographic QCD. We compare the resummed higher-twist cross
sections with the ones obtained in the framework of the frozen
coupling approach and leading-twist cross section. Also, we show
that in the context of frozen coupling approach a higher-twist
contribution to the photon-photon collisions cross section is
normalized in terms of the pion electromagnetic form factor.
\end{abstract}
 \maketitle

\section{\bf Introduction}

One of the most significant theoretical advances in recent years
has been the application of the AdS/CFT correspondence
~\cite{Maldacena} between string theories defined in five-
dimensional anti-de Sitter (AdS) space-time and conformal field
theories in physical space-time. Quantum chromodynamics (QCD) is
not itself a conformal theory; however there are indications, both
from theory ~\cite{Smekal,Furui} and phenomenology
~\cite{Brodsky,Deur} that the QCD coupling is slowly varying at
small momentum transfer. In addition, one can argue that if the
gluon has a maximum wavelength or an effective mass
~\cite{Cornwall} due to confinement, that gluonic vacuum
polarization corrections and the $\beta $-function must vanish in
the infrared. If there is a conformal window where the QCD
coupling is large and approximately constant and quark masses can
be neglected, then QCD resembles a conformal theory, thus
motivating the application of AdS/QCD to QCD. So, even though QCD
is not conformally invariant, one can use the mathematically
representation of the conformal group in five-dimensional anti-de
Sitter space to construct an analytic first approximation to the
theory. The resulting AdS/QCD model gives accurate predictions for
hadron spectroscopy and a description of the quark structure of
mesons and baryons which has scale invariance and dimensional
counting at short distances, together with color confinement at
large distances.

The hadronic wave function in terms of quark and gluon degrees of
freedoms plays an important role in QCD process predictions. For
example, knowledge of the wave function allows to calculate
distribution amplitudes and structure functions or conversely these
processes can give phenomenological restrictions on the wave
functions.

In References~\cite{Bagger,Bagger1,Baier,Ahmadov1,Ahmadov2} the
higher-twist effects was calculated within the frozen coupling
constant approach. In Ref.~\cite{Brodsky1}, it was noted that in
perturbative QCD (pQCD) calculations, the argument of the running
coupling constant in both  renormalization and factorization scale
$Q^2$ should be taken as equal to the square of the momentum
transfer of a hard gluon in a corresponding Feynman diagram. But
defining in this way, $\alpha_{s}({Q}^2)$ suffers from infrared
singularities.

The contribution of large orders of perturbation theory related to
the socalled renormalons has been investigated by several authors,
using in particular the method of Borel
summation~\cite{Hooft,Mueller,Zakharov,Beneke,Greiner}. In the
case of (QCD), the coefficients of perturbative expansions in the
QCD coupling $\alpha_{s}$ can increase dramatically even at low
orders. This fact together with the apparent freedom in the choice
of renormalization scheme and renormalization scales, limits the
predictive power of perturbative calculations, even in
applications involving large momentum transfer, where $\alpha_{s}$
is effectively small.

Investigation of the infrared renormalon effects in various
inclusive and exclusive processes is one of the most important and
interesting problems in the perturbative QCD. It is known that
infrared renormalons are responsible for factorial growth of
coefficients in perturbative series for the physical quantities.
But, these divergent  series can be resummed by means of the Borel
transformation ~\cite{Hooft} and the principal  value
prescription~\cite{Contopanagos} and effects of infrared renormalons
can be taken into account by a scale-setting procedure
$\alpha_{s}(Q^2)\rightarrow\alpha_{s}(exp(f(Q^2))Q^2)$ at the
one-loop order results. Technically, all-order resummation of
infrared renormalons corresponds to the calculation of the one-loop
Feynman diagrams with the running coupling constant
$\alpha_{s}(-k^2)$ at the vertices identically equivalent to
calculation of the same diagrams with nonzero gluon mass.

In this work, we apply the running coupling approach~\cite{Agaev1}
in order to compute the effects of the infrared renormalons on the
pion production in photon-photon collisions within holographic
QCD. This approach was also employed
previously~\cite{Agaev2,Ahmadov3,Ahmadov4,Ahmadov5,Ahmadov6} to
calculate the inclusive meson production in photon-proton,
proton-proton and photon-photon collisions.

Since experiments examining high-$p_T$ particle production in
two-photon collisions have been improved, it becomes important to
reassess the various contributions which arise in quantum
chromodynamics. Also the experimental measurement of the inclusive
charged pion production cross section  is important for the
photon-photon collisions program at the future International Linear
Collider(ILC).

Therefore, the calculation and  analysis of the higher-twist effects
on the dependence of the pion wave function in single pion
production at photon-photon collisions by the running coupling
approach within holographic QCD are very interesting search points.

In this respect, the contribution of the higher-twist Feynman
diagrams to a single meson production cross section in photon-photon
collisions is computed by using various pion wave functions from
holographic QCD. Also, the leading and resummed higher-twist
contributions are estimated and compared to each other.

We organize the  paper as the follows; In Section \ref{ht}, we
provide some formulae for the calculation of the contributions of
the higher twist  and leading twist diagrams. In Section \ref{ir},
we present some formulae and analysis of the higher twist effects on
the dependence of the pion wave function by the running coupling
constant approach, and in Section \ref{results}, the numerical
results for the cross section and discussion for the dependence of
the cross section on the pion wave functions are presented. Finally,
some concluding remark are stated in Section \ref{conc}.

\section{HIGHER TWIST AND LEADING TWIST CONTRIBUTIONS TO INCLUSIVE REACTIONS}\label{ht}
The higher-twist Feynman diagrams for the pion production in the
photon-photon collision $\gamma\gamma \to MX$ are shown in Fig.1(a).
The amplitude for this subprocess can be found by means of the
Brodsky-Lepage formula~\cite{Lepage2}
\begin{equation}
M(\hat s,\hat
t)=\int_{0}^{1}{dx_1}\int_{0}^{1}dx_2\delta(1-x_1-x_2)\Phi_{M}(x_1,x_2,Q^2)T_{H}(\hat
s,\hat t;x_1,x_2).
\end{equation}
In Eq.(2.1), $T_H$ is  the sum of the graphs contributing to the
hard-scattering part of the subprocess.  The hard-scattering
amplitude $T_{H}(\hat s,\hat t;x_1,x_2)$ depends on a process and
can be obtained in the framework of pQCD and represented as a series
in the QCD running coupling constant $\alpha_{s}(Q^2)$. The
light-cone momentum fractions $x\equiv x_1$, $x_2=1-x$ specify the
fractional momenta carried by quark and antiquark in the Fock state.
As higher-twist subprocess which contribute to $\gamma \gamma \to
\pi X$, we take $\gamma q\to Mq$.

The Mandelstam invariant variables for subprocesses $\gamma q\to Mq$
are defined as
\begin{equation}
\hat s=(p_1+p_{\gamma})^2,\quad \hat t=(p_{\gamma}-p_{M})^2,\quad \hat
u=(p_1-p_{M})^2.
\end{equation}

We have aimed to calculate the pion production cross section and
to fix the differences due to the use of various pion model wave
functions. The asymptotic pion wave
functions~\cite{Brodsky3,Brodsky2} and the
Vega-Schmidt-Branz-Gutsche-Lyubovitskij (VSBGL)~\cite{Vega}wave
function predicted by AdS/QCD, and also the pQCD evolution
~\cite{Lepage1} has the form:
$$
\Phi_{asy}^{hol}(x)=\frac{4}{\sqrt{3}\pi}f_{\pi}\sqrt{x(1-x)},
$$
\begin{equation}
\Phi_{VSBGL}^{hol}(x)=\frac{A_1k_1}{2\pi}\sqrt{x(1-x)}exp\left(-\frac{m^2}{2k_{1}^2x(1-x)}\right),\quad
\Phi_{asy}^{p}(x)=\sqrt{3}f_{\pi}x(1-x)
\end{equation}
where $f_{\pi}=92.4 MeV$ is the pion decay constant.

We now incorporate the higher-twist subprocess $\gamma q\to Mq$ into
the full inclusive cross section. In this subprocess photon and the
pion may be viewed as an effective current striking the incoming
quark line. Therefore, the complete cross section in formal analogy
with deep-inelastic scattering, is written as

\begin{equation}
E\frac{d\sigma}{d^{3}p}(\gamma \gamma\to MX )=\frac {3}{\pi}\sum_{q
\overline{q}}\int_{0}^{1}dx \delta(\hat s+\hat t+\hat u)\hat s
G_{q/{\gamma}}(x,-\hat t)\frac{d\sigma}{d\hat t}(\gamma q \to Mq)+
(t\leftrightarrow u).
\end{equation}

Here $G_{q/\gamma}$ is the per color distribution function for a
quark in a photon. The subprocess cross section for $\pi,\rho_{L}$
and $\rho_{T}$ production is
\begin{equation}
\frac{d\sigma}{d\hat t}(\gamma q\to Mq)=\left\{\begin{array}{cc}
\frac{8\pi^2\alpha_{E}C_{F}}{9}[D(\hat s,\hat u)]^{2}
\frac{1}{\hat{s}^2(-\hat t)}\left[\frac{1}{\hat{s}^2}+
\frac{1}{\hat{u}^2}\right],\,\,\, M=\pi,\rho_{L},\\
\frac{8\pi^2\alpha_{E}C_{F}}{9} \left[D(\hat s,\hat
u)\right]^2\frac{8(-\hat t)}{\hat{s}^4 \hat{u}^2},M=\rho_{T},
\end{array} \right.
\end{equation}
 where
\begin{equation}
D(\hat s,\hat u)=e_1\hat t I_{\pi}(Q_1^2)\alpha_{s}(Q_1^2)+e_2\hat
u I_{\pi }(Q_2^2)\alpha_{s}(Q_2^2),
\end{equation}

\begin{equation}
I_{\pi}(Q^2)=\int_{0}^{1}dx\left[\frac{\Phi_{\pi}(x,Q^2)}{x(1-x)}\right]
\end{equation}
and $Q_{1}^2=\hat s/2,\,\,\,\,Q_{2}^2=-\hat u/2$,\,\, represents the
momentum squared carried by the hard gluon in Fig.1(a), $e_1(e_2)$
is the charge of $q_1(\overline{q}_2)$ and $C_F=\frac{4}{3}$.

The $I_{\pi}$ factors reflect the exclusive form factor of the pion
as is the motivation the arguments of $\alpha_{s}$ and $I_{\pi}$.
Note that the relation between $I_{\pi}$ and the pion form factor
completely fixes the normalization of the higher-twist subprocess.
The full cross section for $\pi$ and $\rho_{L}$ production is given
by
$$
E\frac{d\sigma}{d^{3}p}(\gamma \gamma\to MX )=\frac{s}{s+u}
\sum_{q\overline{q}}G_{q/{\gamma}}(x,-\hat
t)\frac{8\pi\alpha_{E}C_{F}}{3}\frac{[D(\hat s,\hat u)]^2}{{\hat
s}^2(-\hat t)}\left[\frac{1}{{\hat s}^2}+\frac{1}{{\hat
u}^2}\right]+
$$
\begin{equation}
\frac{s}{s+t} \sum_{q\overline{q}}G_{q/{\gamma}}(x,-\hat
u)\frac{8\pi\alpha_{E}C_{F}}{3}\frac{[D(\hat s,\hat t)]^2}{{\hat
s}^2(-\hat u)}\left[\frac{1}{{\hat s}^2}+\frac{1}{{\hat
t}^2}\right] ,
\end{equation}

As seen from Eq.(2.8), the subprocess cross section for longitudinal
$\rho_{L}$ production is very similar to that for $\pi$ production.
We have extracted the following higher-twist subprocesses
contributing to the two covariant cross sections in Eq.(2.5) as
\begin{equation}
\gamma q_{1}\to(q_1\overline{q}_2)q_2 \,\,,\,\,\,
\gamma{q}_{2}\to(q_1\overline{q}_2){q}_2.
\end{equation}

Also from Eq.(2.8), at fixed $p_T$, the cross section falls very
slowly with $s$. Additionally, at fixed $s$, the cross section
decreases as $1/p_{T}^5$, multiplied by a slowly varying
logarithmic function which vanishes at the phase-spase boundary.
Thus, the $p_T$ spectrum is  fairly independent of $s$ except near
the kinematic limit.

One of the important problem in the single  inclusive pion
production in photon-photon collision is the possibility of
normalization of the higher-twist subprocess cross section in terms
of the electromagnetic form factor $F_{\pi}(Q^2)$ of the pion. The
electromagnetic form factor $F_{\pi}(Q^2)$ of pion is given by the
expression
\begin{equation}
 F_{\pi}(Q^2)
 =\int_{0}^{1}\int_{0}^{1}{dx_1}dx_2\Phi_{\pi}^{*}(x_1,x_2,Q^2)T_{H}(x_1,x_2,\alpha_{s}(\lambda Q^2),Q^2)\Phi_{\pi}(x_1,x_2,Q^2).
 \end{equation}
This allows us to completely determine  the $\gamma q\to Mq$  cross
section in terms of the pion form factor, through the relation
\begin{equation}
\frac{Q^2 F_{\pi}(Q^2)}{4 \pi C_{F}\alpha_{s}(\hat
Q^2)}=I_{\pi}^{2}(\hat Q^2)
 \end{equation}

It should be noted from Eq.(2.11) that the form factor contains the
square of $I_{\pi}(\hat Q^2)$. In principle, experimental
measurement of $F_{\pi}(Q^2)$ determines $I_{\pi}$ and hence the
$\gamma q\to Mq$. So one can determine the cross section of $\gamma
\gamma\to MX$ explicitly.

Now we can conclude that in the frozen coupling constant approach
$\pi$  meson production  higher-twist cross section of $\gamma
\gamma\to MX$ is normalized in terms of the pion electromagnetic
form factor.

Extracting the higher-twist corrections to the pion production cross
section and a comparison of higher-twist corrections with
leading-twist contributions are essential problems. The contribution
from the leading-twist subprocess $\gamma\gamma\to q\overline{q}$ is
shown in Fig.1(b). The corresponding  inclusive cross section for
production of a meson $M$ is given by
\begin{equation}
\left[\frac{d\sigma}{d^{3}p }\right]_{\gamma\gamma \to
MX}=\frac{3}{\pi}\sum_{q,\overline{q}}\int_{0}^{1}\frac{dz}{z^2}
\delta(\hat{s}+\hat{t}+\hat{u})\hat{s}D_{q}^{M}(z,-\hat{t})\frac{d\sigma}
{d\hat{t}}(\gamma\gamma\to q\overline{q})
\end{equation}
where
$$
 \hat{s}=s,\,\,\hat{t}=\frac{t}{z}\,\,\,\hat{u}=\frac{u}{z}.
$$
Here $s$, $t$, and $u$ refer to the overall $\gamma\gamma\to MX$
reaction. $D_{q}^{M}(z,-\hat t)$ represents the quark fragmentation
function into a pion containing a quark of the same flavor. For
$\pi^{+}$ production we assume that
$D_{\pi^{+}/u}=D_{\pi^{+}/\overline{d}}$. In the leading-twist
subprocess, pion is indirectly emitted from the quark with
fractional momentum $z$.  The final form for the leading-twist
contribution to the large-$p_{T}$ pion production cross section in
the process $\gamma\gamma\to MX$ is
$$
\Sigma_{M}^{LT}\equiv
E\frac{d\sigma}{d^{3}P}=\frac{3}{\pi}\sum_{q,\overline{q}}\int_{0}^{1}\frac{dz}{z^2}\delta
(\hat{s}+\hat{t}+\hat{u})\hat{s}D_{q}^{M}(z,-\hat{t})\frac{d\sigma}
{d\hat{t}}(\gamma\gamma\to q\overline{q})=
$$
\begin{equation}
\frac{3}{\pi}\sum_{q,\overline{q}}\int_{0}^{1}d\frac{1}{z}\delta(s+\frac{1}{z}
(t+u))\hat{s}D_{q}^{M}(z,-\hat{t})\frac{d\sigma}{d\hat{t}}(\gamma\gamma\to
q\overline{q})=
\frac{34}{27}\alpha_{E}^2\frac{1}{z}D_{q}^{M}(z)\frac{1}{{\hat
s}^2}\left[\frac{\hat t}{\hat u}+\frac{\hat u}{\hat t}\right]
\end{equation}

The contributions from these leading-twist subprocesses strongly
depend on some phenomenological factors, such as, quark and gluon
distribution functions in meson and fragmentation functions of
various constituents \emph{etc}. We should note that $D(z,-\hat
t)/z$ behaves as $1/z^2$ when $z\rightarrow0$. For the kinematic
range considered in our numerical calculations, $D(z,-\hat t)/z$
increases even more rapidly. We obtain of the final cross section,
Eq.(2.13), as follows: At fixed $p_T$, the cross section decreases
with $s$ asymptotically as $1/s$. At fixed $s$, the $D(z,-\hat t)$
function causes the cross section to decrease rapidly when $p_T$
increases towards the phase-spase boundary $(z\rightarrow1)$. As
$s$ increases, the phase-spase boundary  moves to higher $p_T$,
and the $p_T$ distribution broadens.

\section{HIGHER TWIST EFFECTS WITHIN HOLOGRAPHIC QCD AND THE ROLE INFRARED RENORMALONS}\label{ir}

The main problem in our investigation is the calculation of the
integral in Eq.(2.6) by the running coupling constant approach
within holographic QCD. It should be noted that, in the exclusive
processes, the coupling constant $\alpha_{s}$ runs not only due to
loop integration, but also because of the integration in the
process amplitude over the light-cone momentum fraction of hadron
constituents. In this respect, the exclusive processes  have two
independent sources of power corrections to their characteristics.
One of the loop integration and the second integration over the
light-cone momentum fraction of hadron constituents. Therefore, it
is worth noting that, the renormalization scale (argument of
$\alpha_s$) according to Fig.1(a) should be chosen to be equal to
$Q_{1}^2=(1-x)\hat s$, and $Q_{2}^2=-x\hat u$. The integral in
Eq.(2.6) takes the form in the framework of the running coupling
approach
\begin{equation}
D(\mu_{R_{0}}^2)=\int_{0}^{1}\frac{\alpha_{s}(\lambda
\mu_{R_0}^2)\Phi_{M}(x,\mu_{F}^2)dx}{x(1-x)}.
\end{equation}
The $\alpha_{s}(\lambda \mu_{R_0}^2)$ has the infrared singularity
at $x\rightarrow1$, for $\lambda=1-x$  or $x\rightarrow0$, for
$\lambda=x$ and so the integral $(3.1)$ diverges. For the
regularization of the integral, we express the running coupling at
scaling variable $\alpha_{s}(\lambda \mu_{R_0}^2)$ with the aid of
the renormalization group equation in terms of the fixed one
$\alpha_{s}(Q^2)$. The solution of renormalization group equation
for the running coupling $\alpha\equiv\alpha_{s}/\pi$ has the form
~\cite{Contopanagos}
\begin{equation}
\frac{\alpha(\lambda)}{\alpha}=\left[1+\alpha
\frac{\beta_{0}}{4}\ln{\lambda}\right]^{-1}.
\end{equation}
Then, for $\alpha_{s}(\lambda Q^2)$, we get
\begin{equation}
\alpha(\lambda Q^2)=\frac{\alpha_{s}}{1+\ln{\lambda/t}}
\end{equation}
where $t=4\pi/\alpha_{s}(Q^2)\beta_{0}=4/\alpha\beta_{0}$.

Having inserted  Eq.(3.3) into Eq.(2.6) we obtain
%\begin{equation}
$$
D(\hat s,\hat u)=e_{1}\hat u\int_{0}^{1}dx\frac{\alpha_{s}(\lambda
\mu_{R_0}^2)\Phi_{M}(x,Q_{1}^2)}{x(1-x)}+ e_{2}\hat
s\int_{0}^{1}dx\frac{\alpha_{s}(\lambda
\mu_{R_0}^2)\Phi_{M}(x,Q_{2}^2)}{x(1-x)}
$$
\begin{equation}
=e_{1}\hat u\alpha_{s}(\hat s)t_{1}\int_{0}^{1}dx
\frac{\Phi_{M}(x,Q_{1}^2)}{x(1-x)(t_{1}+\ln\lambda)} + e_{2}\hat
s\alpha_{s}(-\hat u)t_{2}\int_{0}^{1}dx
\frac{\Phi_{M}(x,Q_{2}^2)}{x(1-x)(t_{2}+\ln\lambda)}
\end{equation}
where $t_1=4\pi/\alpha_{s}(\hat s)\beta_{0}$ and
$t_2=4\pi/\alpha_{s}(-\hat u)\beta_{0}$.

Although the integral (3.4) is still divergent, it is recast into a
suitable  form for calculation. Making the change of variable as
$z=\ln\lambda$, we obtain
\begin{equation}
D(\hat s,\hat u)=e_{1}\hat u \alpha_{s}(\hat s) t_1 \int_{0}^{1}
\frac{\Phi_{M}(x,Q^2)dx}{x(1-x)(t_{1}+z)}+
%\begin{equation}
e_{2}\hat s \alpha_{s}(-\hat u) t_2 \int_{0}^{1}
\frac{\Phi_{M}(x,Q^2)dx}{x(1-x)(t_{2}+z)}
\end{equation}
In order to calculate (3.5), we will apply the integral
representation of $1/(t+z)$ ~\cite{Zinn-Justin,Erdelyi} as
\begin{equation}
\frac{1}{t+z}=\int_{0}^{\infty}e^{-(t+z)u}du
\end{equation}
which gives
\begin{equation}
D(\hat s,\hat u)=e_{1} \hat{t} \alpha_{s}(\hat s) t_1 \int_{0}^{1}
\int_{0}^{\infty} \frac{\Phi_{\pi}(x,Q_{1}^2)e^{-(t_1+z)u}du
dx}{x(1-x)}+ e_{2} \hat{u} \alpha_{s}(-\hat u) t_2 \int_{0}^{1}
\int_{0}^{\infty} \frac{\Phi_{\pi}(x,Q_{2}^2)e^{-(t_2+z)u}du
dx}{x(1-x)}.
\end{equation}
In the case $\Phi_{asy}^{hol}(x)$ for the $D(\hat s,\hat u)$,
Eq.(3.7) is written as
%\begin{equation}
\begin{equation}
D(\hat s,\hat u)=\frac{16 f_{\pi} e_{1} \hat u}{\sqrt{3}\beta_{0}}
\int_{0}^{\infty} du
e^{-t_{1}u}B\left(\frac{1}{2},\frac{1}{2}-u\right)+  \frac{16
f_{\pi} e_{2} \hat s}{\sqrt{3}\beta_{0}} \int_{0}^{\infty} du
e^{-t_{2}u}B\left(\frac{1}{2},\frac{1}{2}-u\right)
\end{equation}
and for $\Phi_{asy}^{p}(x)$ wave function
%\begin{equation}
\begin{equation}
D(\hat s,\hat u)=\frac{4\sqrt{3}\pi f_{\pi}e_{1}\hat u}{\beta_{0}}
\int_{0}^{\infty}du e^{-t_{1}u} \left[\frac{1}{1-u}\right]
+\frac{4\sqrt{3}\pi f_{\pi}e_{2}\hat s}{\beta_{0}}
\int_{0}^{\infty}du e^{-t_{2}u} \left[\frac{1}{1-u}\right].
\end{equation}
where $B(\alpha,\beta)$ is Beta function.  The structure of the
infrared renormalon poles in Eq.(3.8) and Eq.(3.9) strongly depend
on the wave functions of the pion. To remove them from Eq.(3.8) and
Eq.(3.9) we adopt the principal value prescription. We denote the
higher-twist cross section obtained using the running coupling
constant approach by $(\Sigma_{\pi}^{HT})^{res}$.

\section{NUMERICAL RESULTS AND DISCUSSION}\label{results}

In this section, we discuss the numerical results for higher-twist
and renormalon effects with higher-twist contributions calculated
in the context of the running coupling constant  and frozen
coupling approaches on the dependence of the chosen pion wave
functions in the process $\gamma \gamma \to MX$ within holographic
QCD. For the higher-twist subprocess, we take $\gamma
q_{1}\to(q_1\overline{q}_2)q_2$, and $\gamma
q_{2}\to(q_1\overline{q}_2){q}_2$ contributing to $\gamma\gamma\to
MX$ cross sections. Inclusive pion photoproduction represents a
significant test case in which higher-twist terms dominate those
of leading-twist terms in certain kinematic domains. For the
dominant leading-twist subprocess for the meson production, we
take the photon-photon annihilation $\gamma\gamma \to q\bar{q}$ in
which the $\pi$ pion is indirectly emitted from the quark. The
quark distribution function inside the photon was used
~\cite{Cornet} and the gluon and quark fragmentation functions
into a pion was used ~\cite{Albino}.

The results of our numerical calculations are plotted in
Figs.2-15. Firstly, it is very interesting comparing the higher
twist cross sections obtained within  holographic QCD with the
ones obtained within perturbative QCD. In Fig.2 and Fig.3 we show
the dependence of higher-twist cross sections
$(\Sigma_{\pi^{+}}^{HT})^{0}$, and $(\Sigma_{\pi^{+}}^{HT})^{res}$
calculated in the context of the frozen and  running coupling
constant approaches as a function of the pion transverse momentum
$p_{T}$ for $\Phi_{\pi}^{hol}(x)$, $\Phi_{\pi}^{p}(x)$ and
$\Phi_{VSBGL}^{hol}(x)$ pion wave functions at $y=0$. It is seen
from the figures that the higher-twist cross section is
monotonically decreasing with an increase in the transverse
momentum of the pion. In Fig.4-Fig.7, we show the dependence of
the ratios
$(\Sigma_{HT}^{hol})^{0}$/$(\Sigma_{\pi^{+}}^{HT})^{0}$,
$(\Sigma_{\pi^{+}}^{HT})^{res}$/$(\Sigma_{\pi^{+}}^{HT})^{0}$,
$(\Sigma_{\pi^{+}}^{HT})^{0}$/$(\Sigma_{\pi^{+}}^{LT})$ and
$(\Sigma_{\pi^{+}}^{HT})^{res}$/$(\Sigma_{\pi^{+}}^{LT})$ as a
function of the pion transverse momentum $p_{T}$ for
$\Phi_{\pi}^{hol}(x)$, $\Phi_{\pi}^{p}(x)$ and
$\Phi_{VSBGL}^{hol}(x)$ pion wave functions. Here
$\Sigma_{\pi^{+}}^{LT}$ is the leading-twist cross section. As
shown in Fig.4, in the region $20\,\,GeV/c<p_T<80\,\,GeV/c$
higher-twist cross section for $\Phi_{\pi}^{hol}(x)$ is suppress
by about half orders of magnitude relative to the  higher-twist
cross section for $\Phi_{VSBGL}^{hol}(x)$, but in the regions
$10\,\,GeV/c<p_T<20\,\,GeV/c$ and $80\,\,GeV/c<p_T<90\,\,GeV/c$,
higher-twist cross section $(\Sigma_{HT}^{hol})^{0}$ is suppress
by about two orders of magnitude relative to the  higher-twist
cross section for $(\Sigma_{VSBGL}^{hol})^{0}$. Also higher-twist
cross section for $\Phi_{\pi}^{p}(x)$ is suppress by about half
orders of magnitude relative to the higher-twist cross section for
$\Phi_{VSBGL}^{hol}(x)$ . In Fig.5, Fig.6 and Fig.7, the
dependence of the ratios
$(\Sigma_{\pi}^{HT})^{res}$/$(\Sigma_{\pi^{+}}^{HT})^{0}$,
$(\Sigma_{\pi^{+}}^{HT})^{0}$/$(\Sigma_{\pi^{+}}^{LT})$ and
$(\Sigma_{\pi^{+}}^{HT})^{res}$/$(\Sigma_{\pi^{+}}^{LT})$ are
displyed as a function of the pion transverse momentum $p_{T}$ for
the $\Phi_{\pi}^{hol}(x)$,  $\Phi_{\pi}^{p}(x)$ and
$\Phi_{VSBGL}^{hol}(x)$ pion wave functions. It is see that,
resummed higher-twist cross section for $\Phi_{asy}^{hol}(x)$ is
suppress by about two orders of magnitude relative to the
higher-twist cross section for $(\Sigma_{VSBGL}^{hol})^{0}$.
Noticed that one-half order is suppress for
$(\Sigma_{asy}^{hol})^{0}$ and one order is suppress for
$(\Sigma_{asy}^{p})^{0}$. It is observed from Fig.6 and Fig.7
that, the ratios
$(\Sigma_{\pi^{+}}^{HT})^{0}$/$(\Sigma_{\pi^{+}}^{LT})$ and
$(\Sigma_{\pi^{+}}^{HT})^{res}$/$(\Sigma_{\pi^{+}}^{LT})$  for all
wave functions decrease with an increase in the transverse
momentum of pion. In Fig.8 - Fig.10, we have depicted higher-twist
cross sections $(\Sigma_{HT}^{hol})^{0}$, and ratio
$(\Sigma_{HT}^{hol})^{0}/(\Sigma_{\pi^{+}}^{HT})^{0}$,
$(\Sigma_{HT}^{hol})^{res}/(\Sigma_{\pi^{+}}^{HT})^{0}$, as a
function of the rapidity $y$ of the pion at $\sqrt s=183\,\,GeV$
and $p_T=14.6 \,\,GeV/c$. The figures show that the higher-twist
cross section and ratios have a different distinctive behavior. As
is seen in Fig.9 ratio
$(\Sigma_{asy}^{hol})^{0}/(\Sigma_{VSBGL}^{hol})^{0}$ has a
maximum approximately at the point $y=-1.92$. However in this
point ratio $(\Sigma_{VSBGL}^{hol})^{0}/(\Sigma_{asy}^{p})^{0}$
has a minimum. As is seen from Fig.10 resummed higher-twist cross
section  for $\Phi_{asy}^{hol}(x)$ is suppress by about one  order
of magnitude relative to the resummed higher-twist cross section
for $\Phi_{asy}^{p}(x)$ and with an increasing  rapidity  of pion
ratio is kepped  approximately constant. But resummed higher-twist
cross section  for $\Phi_{asy}^{hol}(x)$ is suppress by about one
half- two orders of magnitude relative to the frozen  higher-twist
cross section for $\Phi_{VSBGL}^{hol}(x)$ and  has a maximum
approximately at the point $y=-1.92$. However resummed
higher-twist cross section for $\Phi_{asy}^{hol}(x)$ is suppress
by about one order of magnitude relative for resummed higher-twist
cross section for $\Phi_{asy}^{p}(x)$ and to stay is constant with
an increasing rapidity of pion. Figures also show that, the ratio
depends on the choice of the pion wave function. Analysis of our
calculations concludes that $(\Sigma_{\pi^{+}}^{HT})^{0}$, and
$(\Sigma_{\pi^{+}}^{HT})^{res}$ higher-twist cross sections and
ratio sensitive to pion wave functions predicted by holographic
and perturbative QCD.

We have also carried out comparative calculations in the
center-of-mass energy $\sqrt s=209\,\,GeV$ and obtained results
are displayed in Fig.11-Fig.15.  Analysis of our calculations at
the center-of-mass energies $\sqrt s=183\,\,GeV$ and $\sqrt
s=209\,\,GeV$, show that with increasing in the beam energy
contributions of higher twist effects  to the cross section
decrease by about 1-2 orders of magnitude. As is seen from Fig.4,
Fig.5, Fig.9, Fig.10, Fig.12, Fig.14 and Fig.15 that infrared
renormalon effects enhance the perturbative predictions for the
pion production cross section in the photon-photon collisions
about 1-2 orders of magnitude. Our opinion is this feature of
infrared renormalons may help the explain theoretical
interpretations with future experimental data for the pion
production cross section in the photon-photon collisions. In our
calculations of the higher-twist cross section of the process the
dependence of the transverse momentum of pion appears in the range
of $(10^{-9}\div10^{-22})mb/GeV^2$. Therefore, higher-twist cross
section obtained in our work should be observable at ILC.

\section{CONCLUSIONS}\label{conc}
In this work, the single meson inclusive production via higher
twist mechanism within holographic QCD are calculated . For
calculation of the cross section  the running coupling constant
approach is applied and infrared renormalon poles in the cross
section expression are revealed. Infrared renormalon induced
divergences is regularized by means of the principal value
prescripton and the Borel sum for the higher twist cross section
is find. It is observed that, the resummed higher-twist cross
section differs from that found using the frozen coupling
approximation in some region considerably. We proved that the
higher-twist cross section for $\pi$ pion production in the
photon-photon collisions  may be normalized in terms of the pion
form factor. The following results can be concluded from the
experiments; the higher-twist contributions to single meson
production cross section in the photon-photon collisions have
important phenomenological consequences, the higher-twist pion
production cross section in the photon-photon collisions depends
on the form of the pion model wave functions and may be used for
their study. Also that the contributions of renormalons effects
within holograpich QCD in this process is essential and may  help
to analyse  experimental results. Further investigations are
needed in order to clarify the role of higher-twist effects  in
QCD. Especially, the future ILC measurements will provide further
tests of the dynamics of large-$p_T$ hadron production beyond the
leading twist.

\section*{Acknowledgments}
One of author A.I.Ahmadov is grateful to all members of the
Department of Physics of Karadeniz Technical University for
appreciates hospitality extended to him in Trabzon. Financial
support by TUBITAK under grant number 2221(Turkey) is also
gratefully acknowledged.

\newpage

%\begin{center}
%{\bf FIGURE CAPTIONS \\}
%\end{center}
%\noindent
\begin{figure}[!hbt]
\epsfxsize 8cm \centerline{\epsfbox{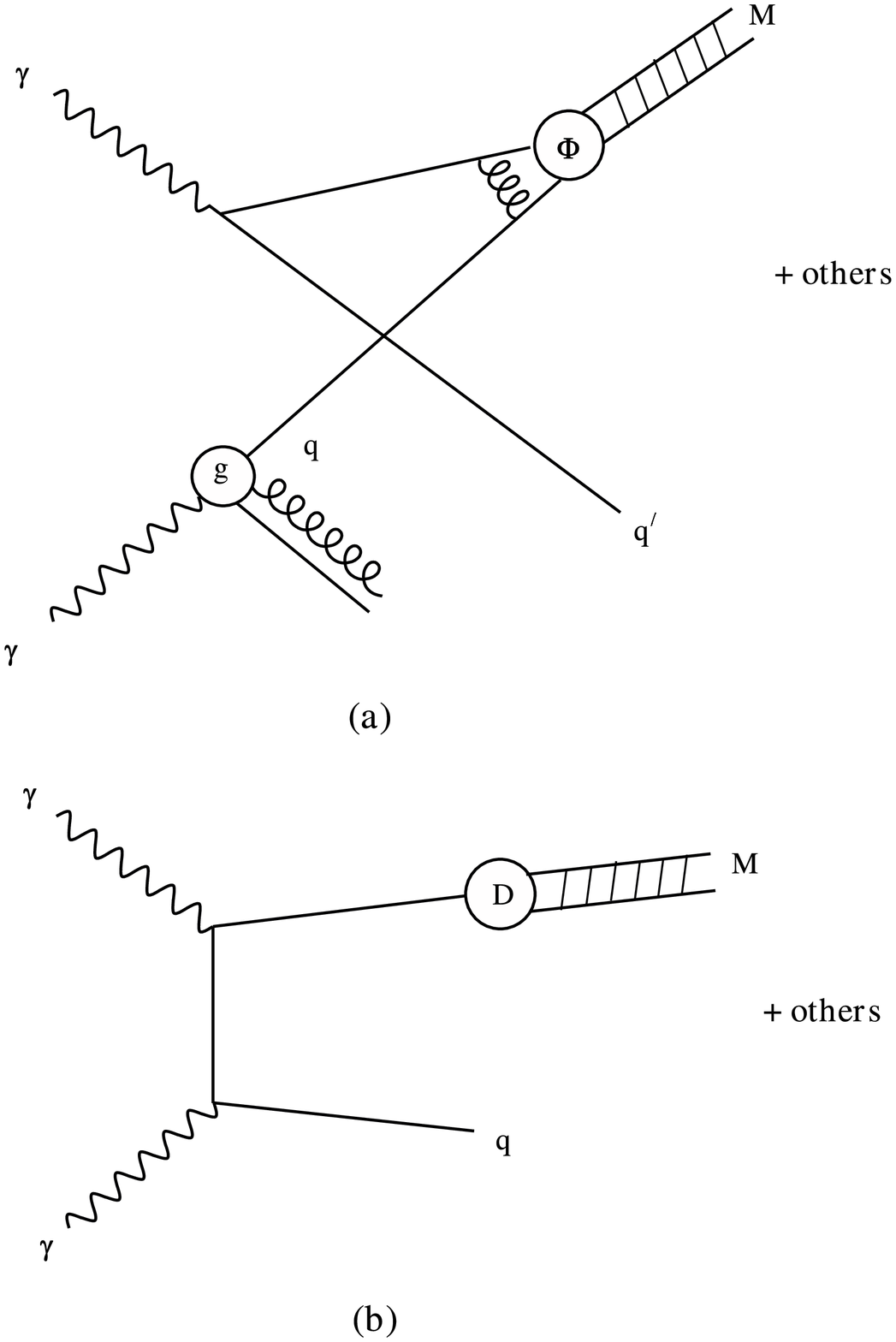}} \vskip -0.02cm
\caption{(a): The higher-twist contribution to $\gamma\gamma\to
MX$;\,\,\,(b): The leading-twist contribution to  $\gamma\gamma\to
MX$} \label{Fig1}
\end{figure}

\begin{figure}[!hbt]
\vskip -1.2cm\epsfxsize 11.8cm \centerline{\epsfbox{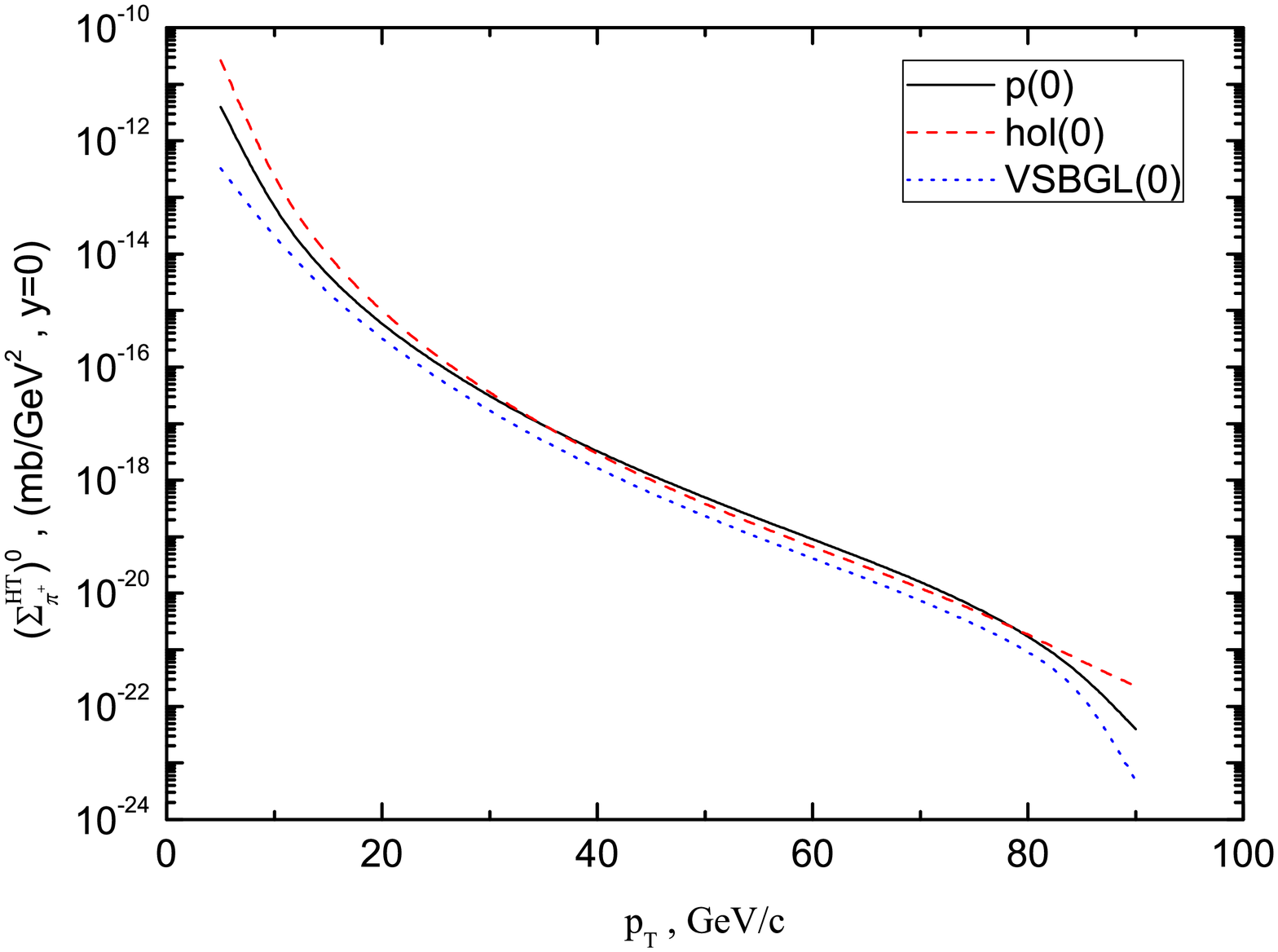}}
\vskip-0.2cm \caption{Higher-twist $\pi^{+}$ production cross
section $(\Sigma_{\pi^{+}}^{HT})^{0}$ as a function of the $p_{T}$
transverse momentum of the pion at the c.m. energy $\sqrt s=183\,\,
GeV$.} \label{Fig2}
\end{figure}

\begin{figure}[!hbt]
\vskip 1.2cm \epsfxsize 11.8cm \centerline{\epsfbox{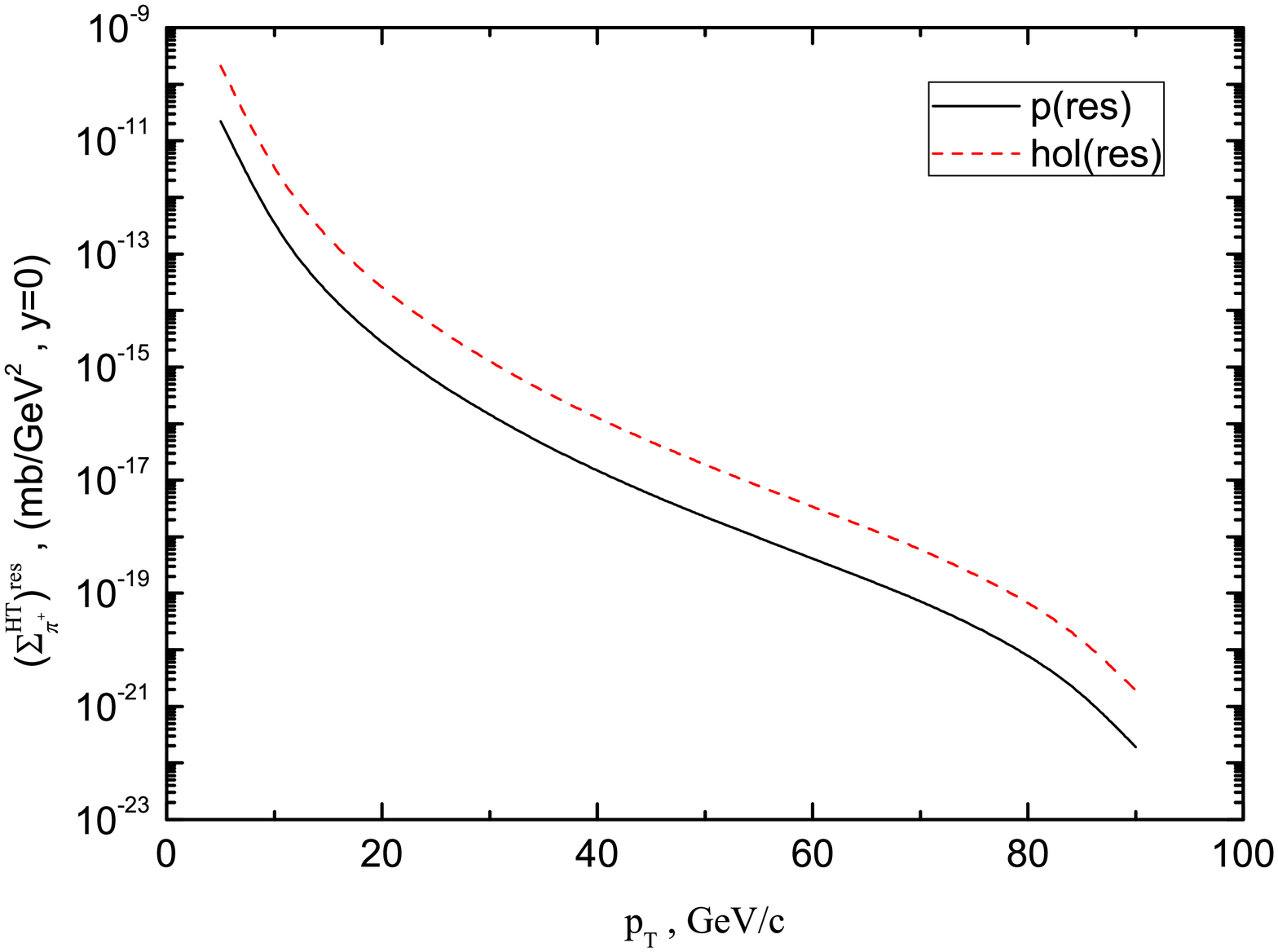}}
\vskip-0.2cm \caption{Higher-twist $\pi^{+}$ production cross
section $(\Sigma_{\pi^{+}}^{HT})^{res}$ as a function of the $p_{T}$
transverse momentum of the pion at the c.m.energy $\sqrt s=183\,\,
GeV$.} \label{Fig3}
\end{figure}

\begin{figure}[!hbt]
\vskip -1.2cm\epsfxsize 11.8cm \centerline{\epsfbox{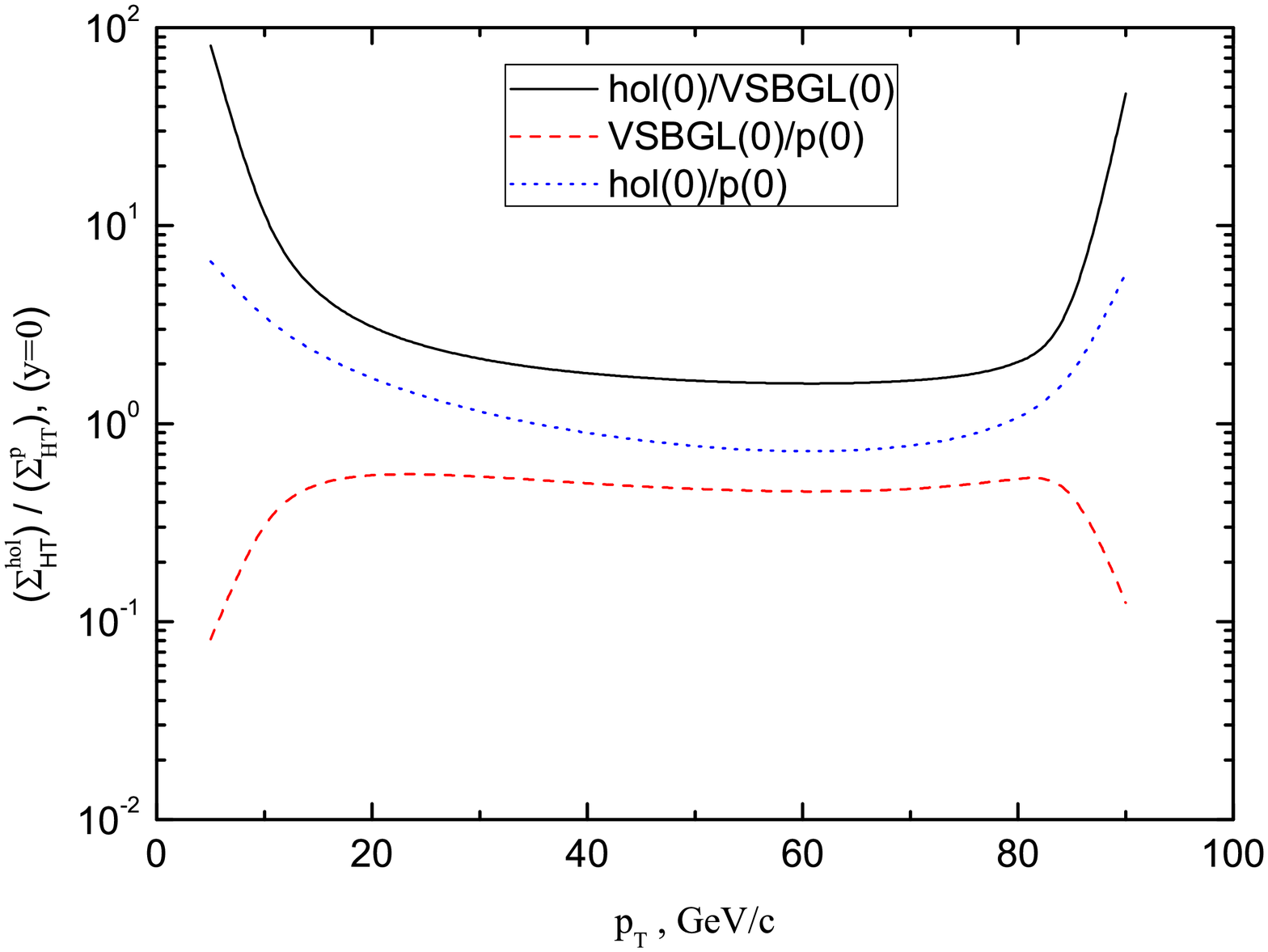}}
\vskip-0.2cm \caption{Ratio
$(\Sigma_{HT}^{hol})^{0}/(\Sigma_{\pi^{+}}^{HT})^{0}$, where
higher-twist contribution are calculated for the pion rapidity $y=0$
at the c.m.energy $\sqrt s=183\,\, GeV$ as a function of the pion
transverse momentum, $p_{T}$.} \label{Fig4}
% \vskip 1.8cm
\end{figure}

\begin{figure}[!hbt]
\vskip 1.2cm\epsfxsize 11.8cm \centerline{\epsfbox{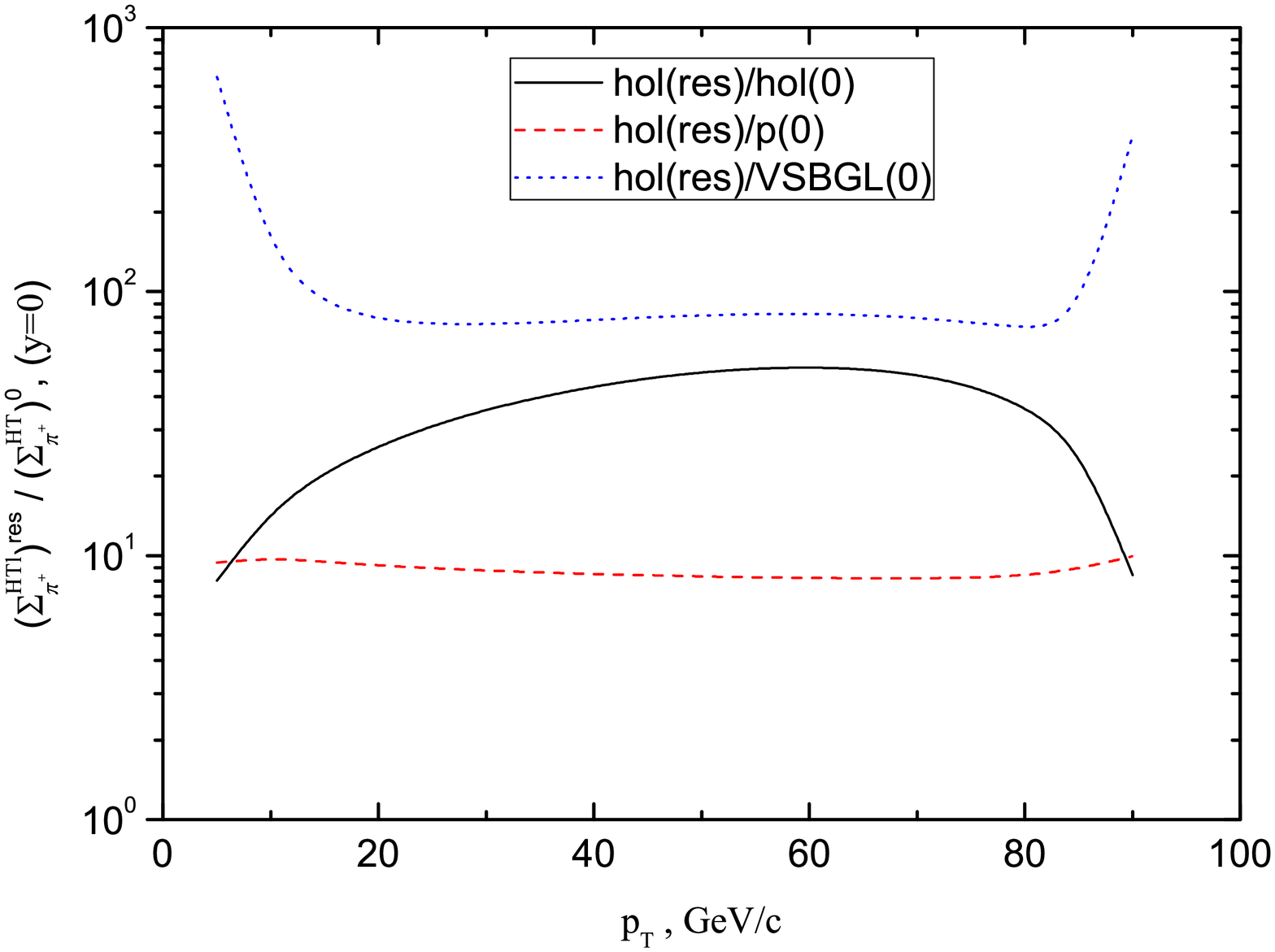}}
\vskip-0.2cm \caption{Ratio
$(\Sigma_{HT}^{hol})^{res}/(\Sigma_{\pi^{+}}^{HT})^0$, as a function
of the $p_{T}$ transverse momentum of the pion at the c.m. energy
$\sqrt s=183\,\, GeV$.} \label{Fig5}
\end{figure}

\begin{figure}[!hbt]
\vskip -1.2cm\epsfxsize 11.8cm \centerline{\epsfbox{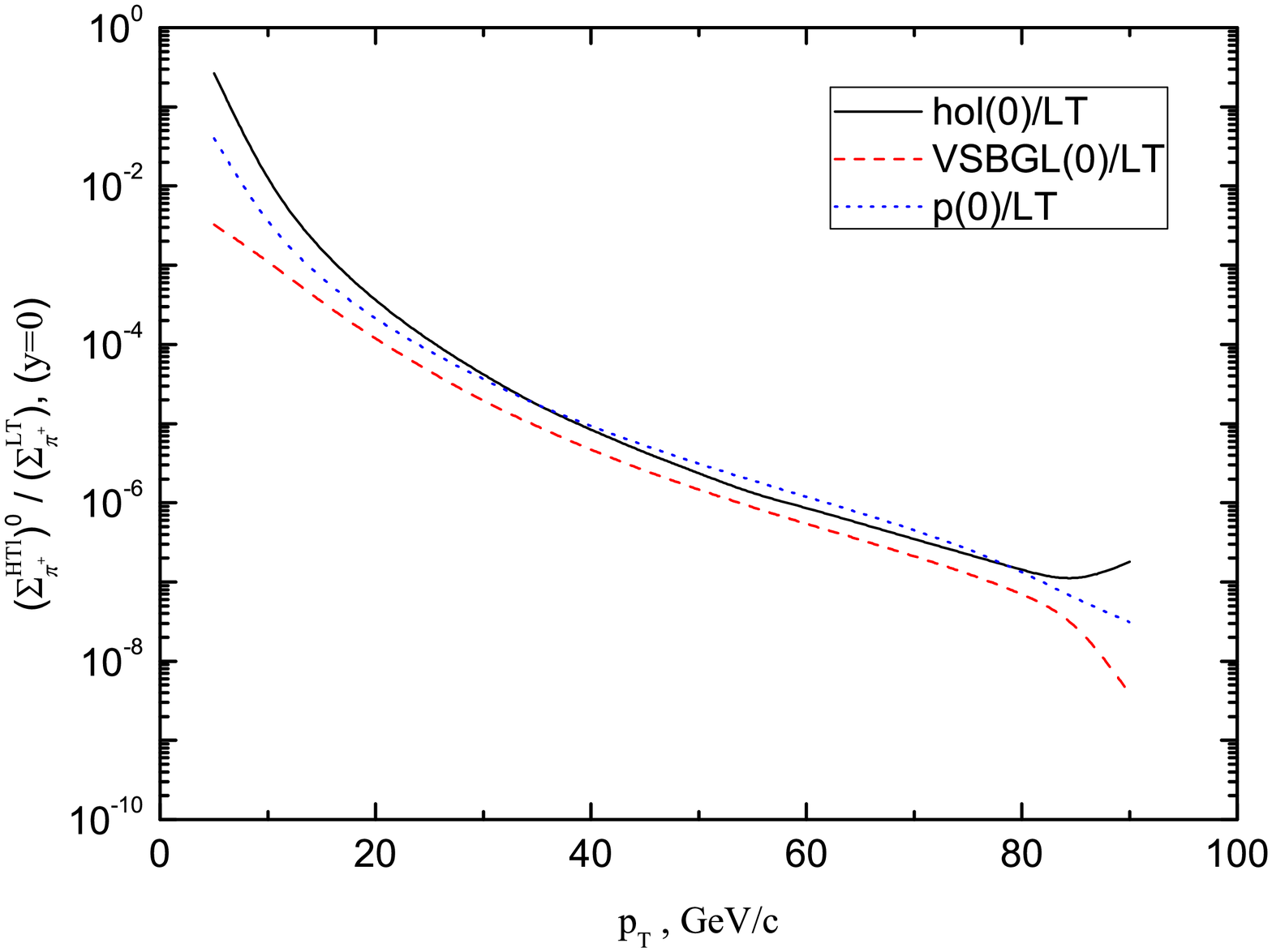}}
\vskip-0.2cm \caption{Ratio
$(\Sigma_{\pi^{+}}^{HT})^0/(\Sigma_{\pi^{+}}^{LT})$, where
higher-twist contribution are calculated for the pion rapidity $y=0$
at the c.m.energy $\sqrt s=183\,\, GeV$ as a function of the pion
transverse momentum, $p_{T}$.} \label{Fig6}
% \vskip 1.8cm
\end{figure}

\begin{figure}[!hbt]
\vskip 1.2cm\epsfxsize 11.8cm \centerline{\epsfbox{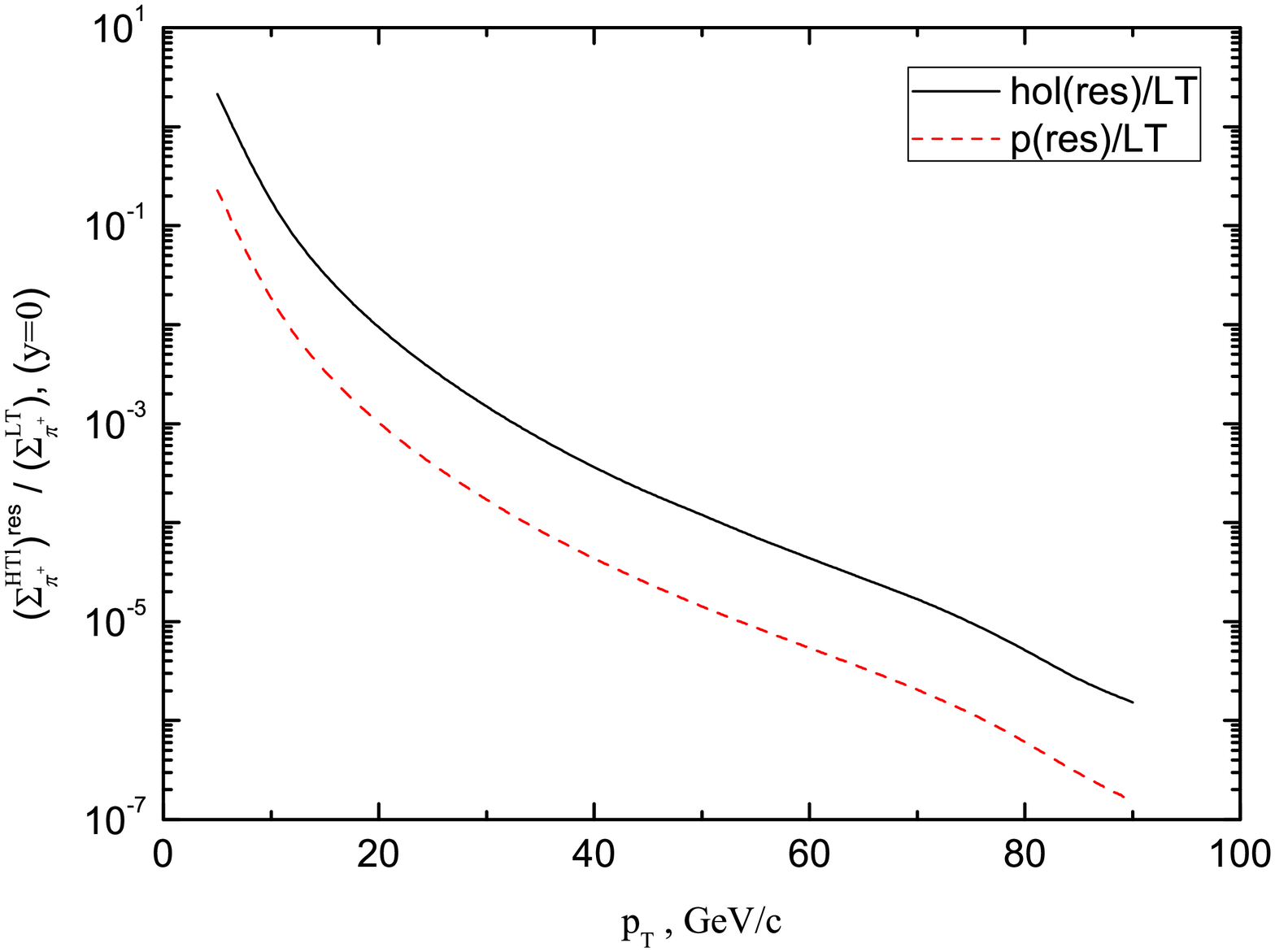}}
\vskip-0.2cm \caption{Ratio
$(\Sigma_{\pi^{+}}^{HT})^{res}/(\Sigma_{\pi^{+}}^{LT})$, as a
function of the $p_{T}$ transverse momentum of the pion at the
c.m. energy $\sqrt s=183\,\, GeV$.} \label{Fig7}
\end{figure}

\begin{figure}[!hbt]
\vskip-1.2cm \epsfxsize 11.8cm \centerline{\epsfbox{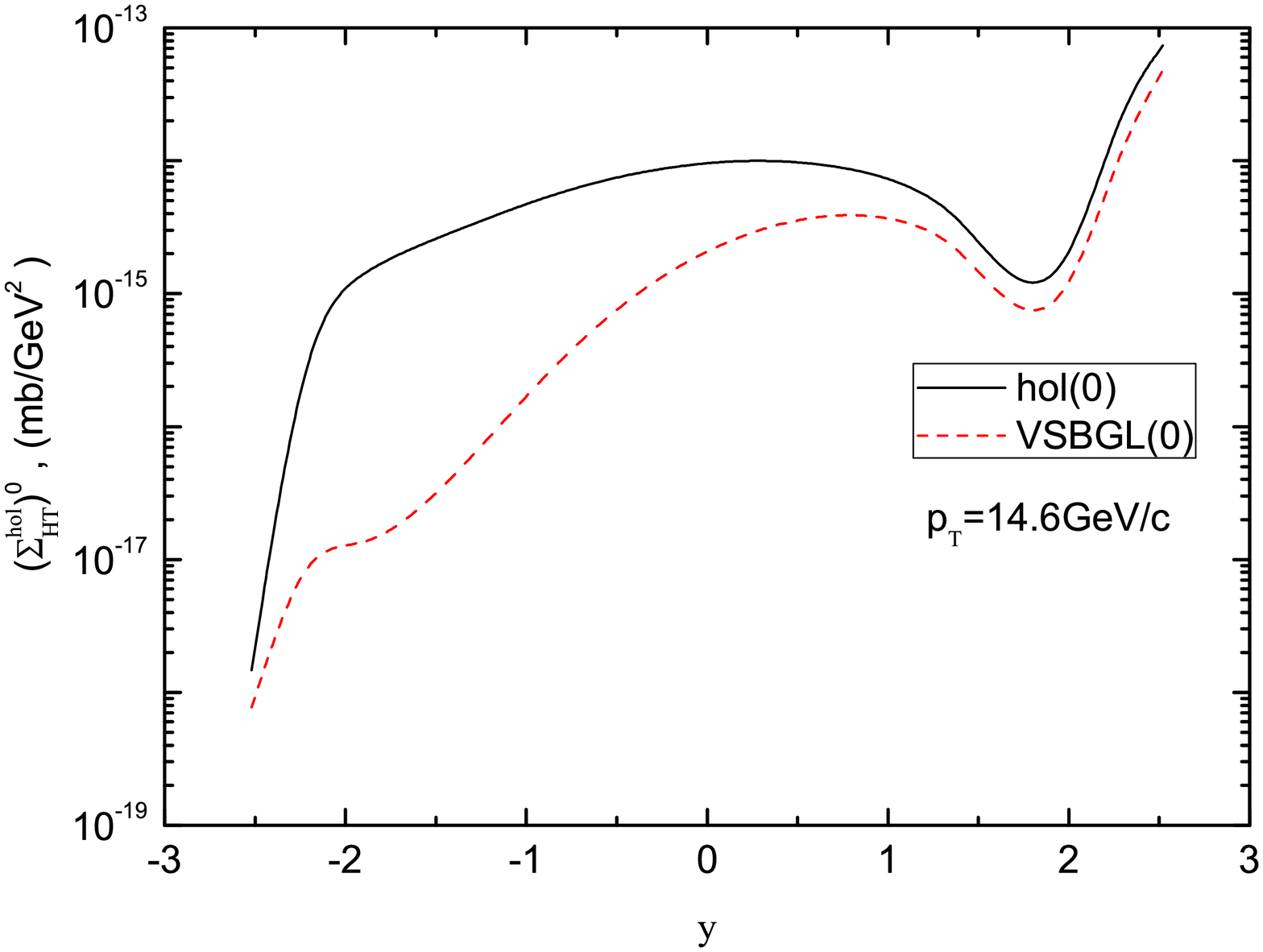}}
\vskip-0.2cm \caption{Higher-twist $\pi^{+}$ production cross
section  $(\Sigma_{HT}^{hol})^0$ , as a function of the $y$ rapidity
of the pion at the  transverse momentum of the pion $p_T=14.6\,\,
GeV/c$, at the c.m. energy $\sqrt s=183\,\, GeV$.} \label{Fig8}
\end{figure}

\begin{figure}[!hbt]
\vskip 0.8cm \epsfxsize 11.8cm \centerline{\epsfbox{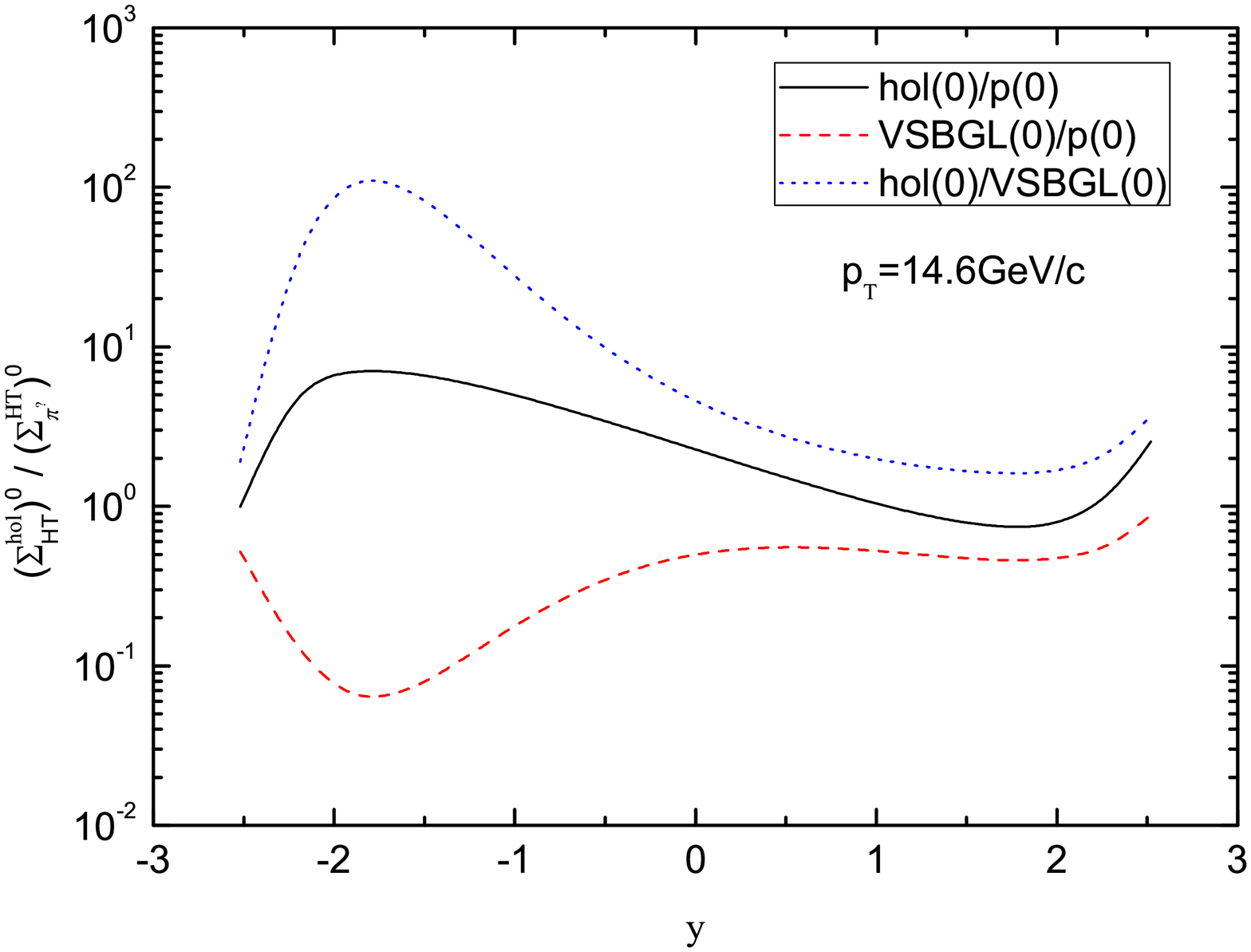}}
\vskip-0.2cm \caption{Ratio
$(\Sigma_{HT}^{hol})^{0}/(\Sigma_{\pi^{+}}^{HT})^0$, as a function
of the $y$ rapidity of the pion at the  transverse momentum of the
pion $p_T=14.6\,\, GeV/c$, at the c.m. energy $\sqrt s=183\,\,
GeV$.} \label{Fig9}
\end{figure}

\begin{figure}[!hbt]
\vskip-1.2cm \epsfxsize 11.8cm \centerline{\epsfbox{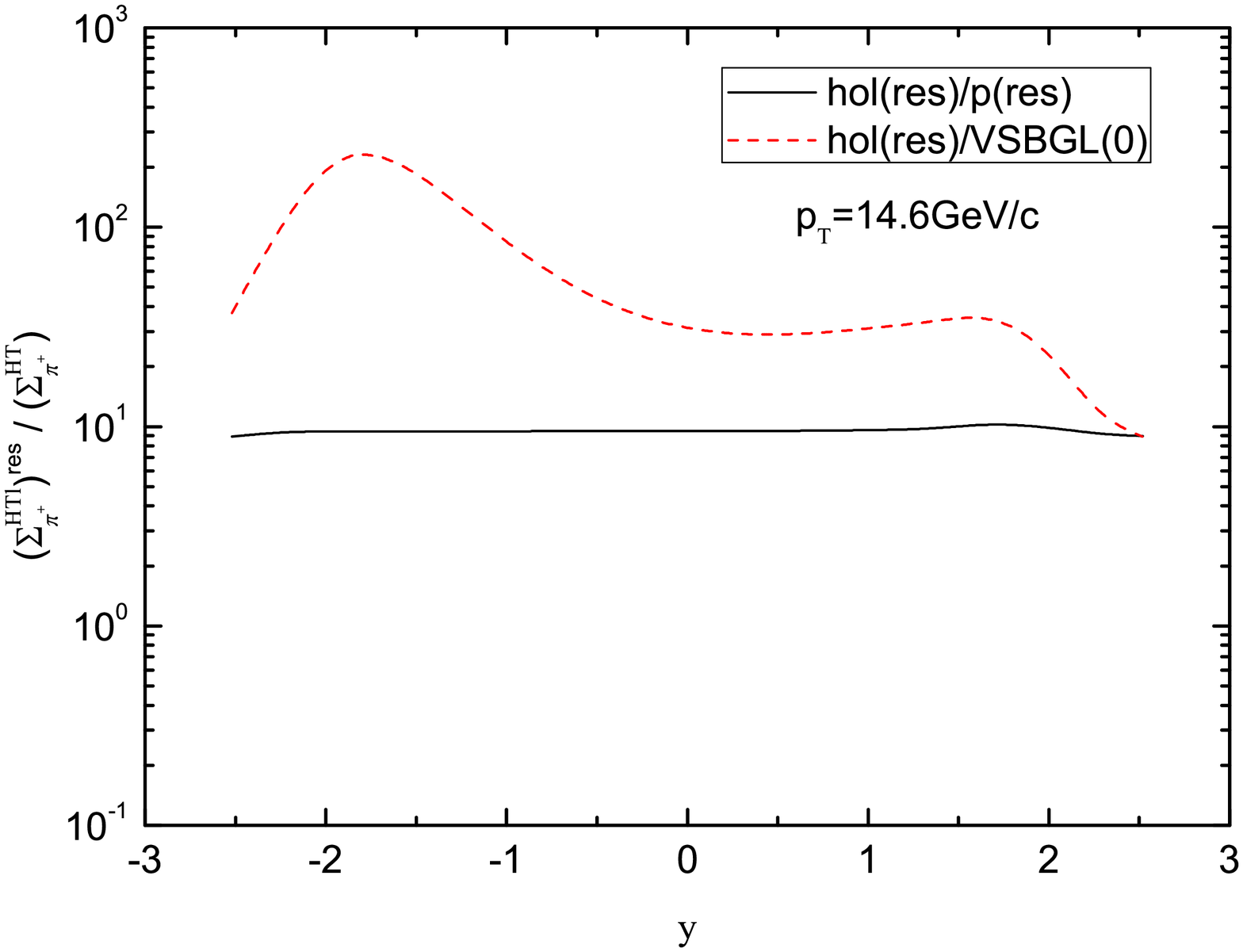}}
\vskip-0.2cm \caption{Ratio
$(\Sigma_{\pi^{+}}^{HT})^{res}/(\Sigma_{\pi^{+}}^{HT})^{0}$, as a
function of the $y$ rapidity of the pion at the  transverse momentum
of the pion $p_T=14.6\,\, GeV/c$, at the c.m. energy $\sqrt
s=183\,\, GeV$.} \label{Fig10}
\end{figure}

\begin{figure}[!hbt]
\vskip-1.2cm\epsfxsize 11.8cm \centerline{\epsfbox{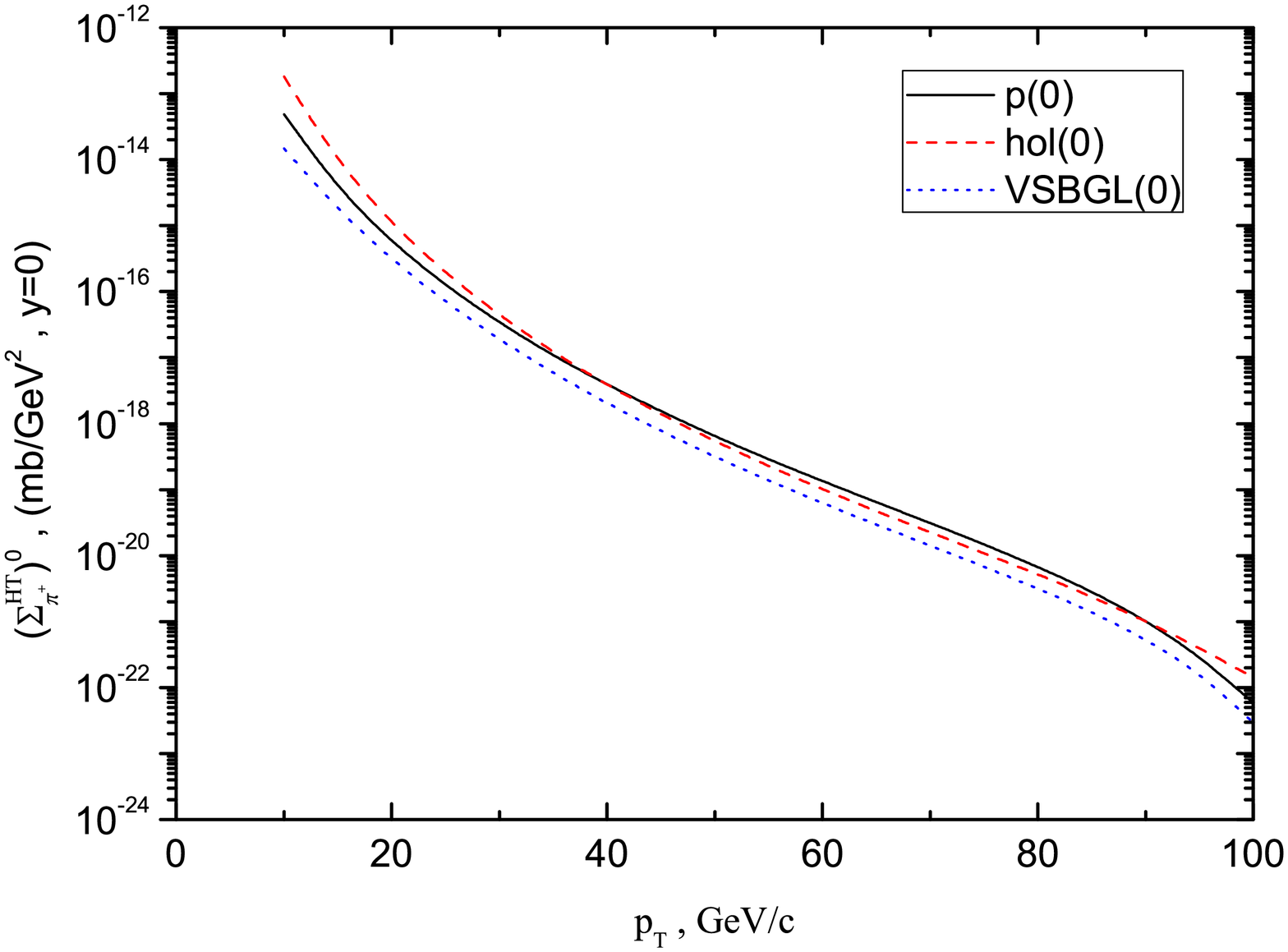}}
\vskip-0.2cm \caption{Higher-twist $\pi^{+}$ production cross
section $(\Sigma_{\pi^{+}}^{HT})^{0}$ as a function of the $p_{T}$
transverse momentum of the pion at the c.m.energy $\sqrt s=209\,\,
GeV$.} \label{Fig11}
\end{figure}

\begin{figure}[!hbt]
\epsfxsize 11.8cm \centerline{\epsfbox{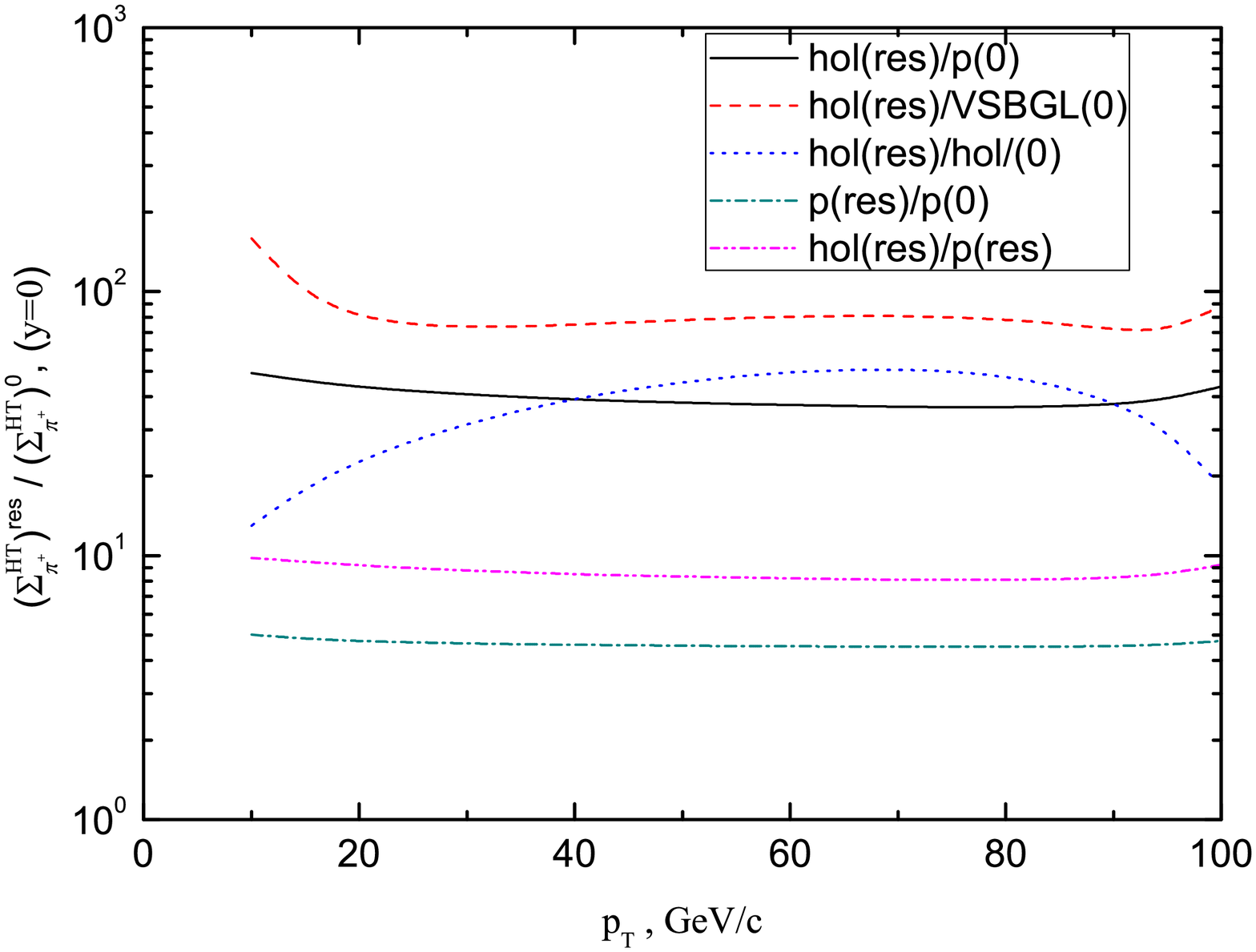}} \vskip-0.05cm
\caption{Ratio
$(\Sigma_{HT}^{hol})^{res}/(\Sigma_{\pi^{+}}^{HT})^{0}$, as a
function of the $p_{T}$ transverse momentum of the pion at the
c.m.energy $\sqrt s=209\,\, GeV$.} \label{Fig12}
\end{figure}

\begin{figure}[!hbt]
\vskip-1.2cm\epsfxsize 11.8cm \centerline{\epsfbox{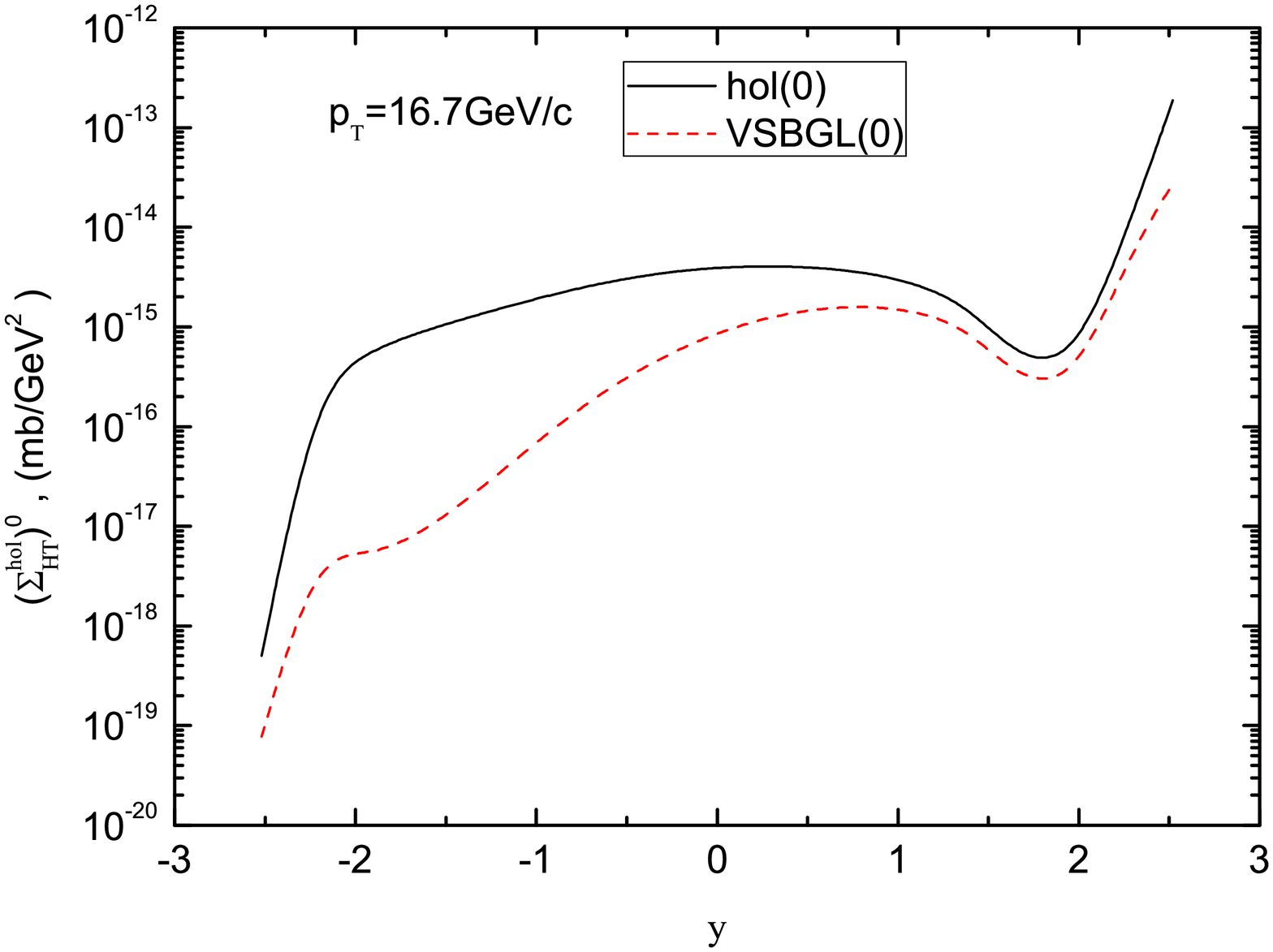}}
\vskip-0.2cm \caption{Higher-twist $\pi^{+}$ production cross
section  $(\Sigma_{HT}^{hol})^0$ , as a function of the $y$ rapidity
of the pion at the  transverse momentum of the pion $p_T=16.7\,\,
GeV/c$, at the c.m. energy $\sqrt s=209\,\, GeV$.} \label{Fig13}
\vskip 1.8cm
\end{figure}

\begin{figure}[!hbt]
\vskip 0.8cm \epsfxsize 11.8cm \centerline{\epsfbox{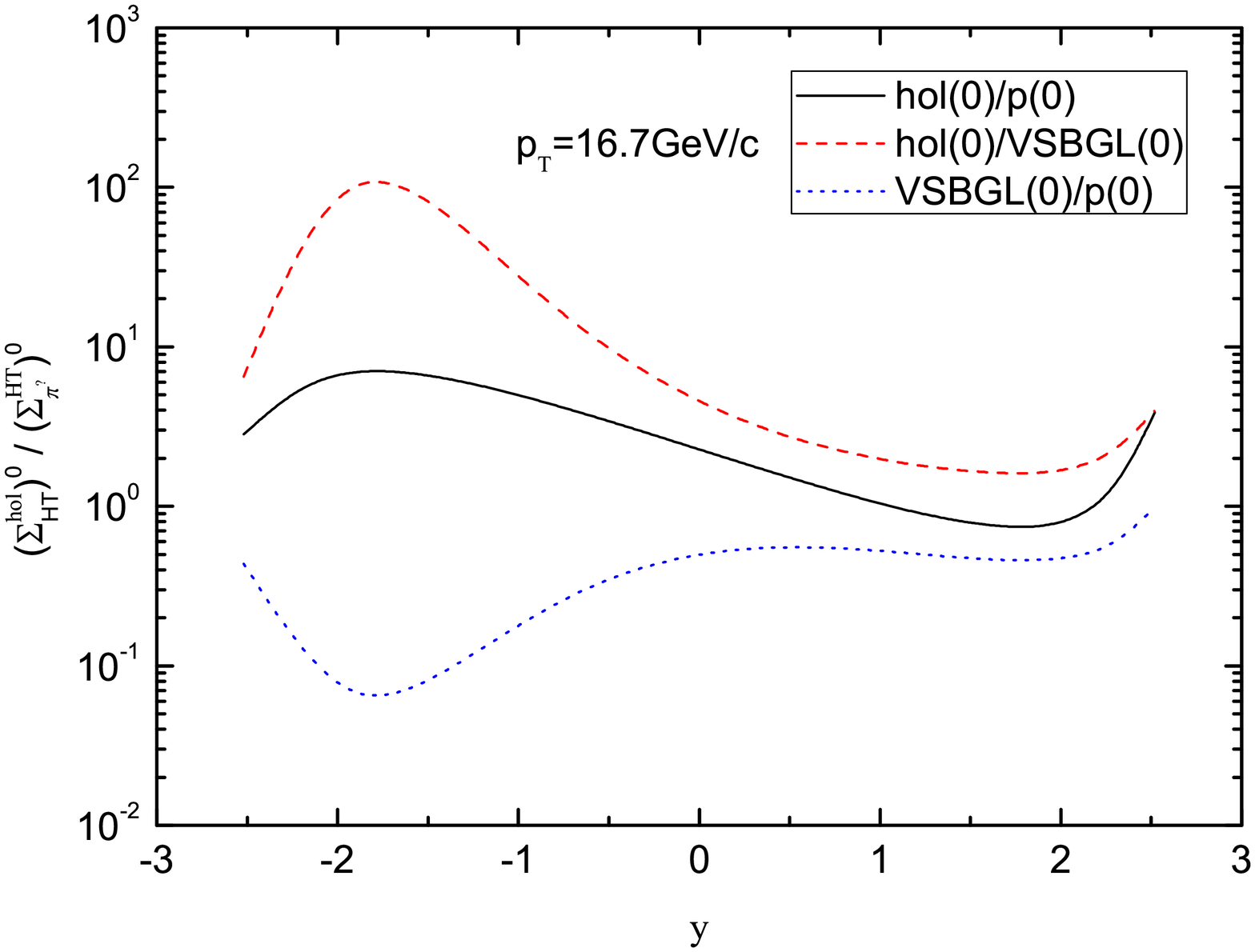}}
\vskip-0.2cm \caption{Ratio
$(\Sigma_{HT}^{hol})^{0}/(\Sigma_{\pi^{+}}^{HT})^{0}$, as a function
of the $y$ rapidity of the pion at the  transverse momentum of the
pion $p_T=16.7\,\, GeV/c$, at the c.m. energy $\sqrt s=209\,\,
GeV$.} \label{Fig14}
\end{figure}

\begin{figure}[!hbt]
\vskip 0.8cm \epsfxsize 11.8cm \centerline{\epsfbox{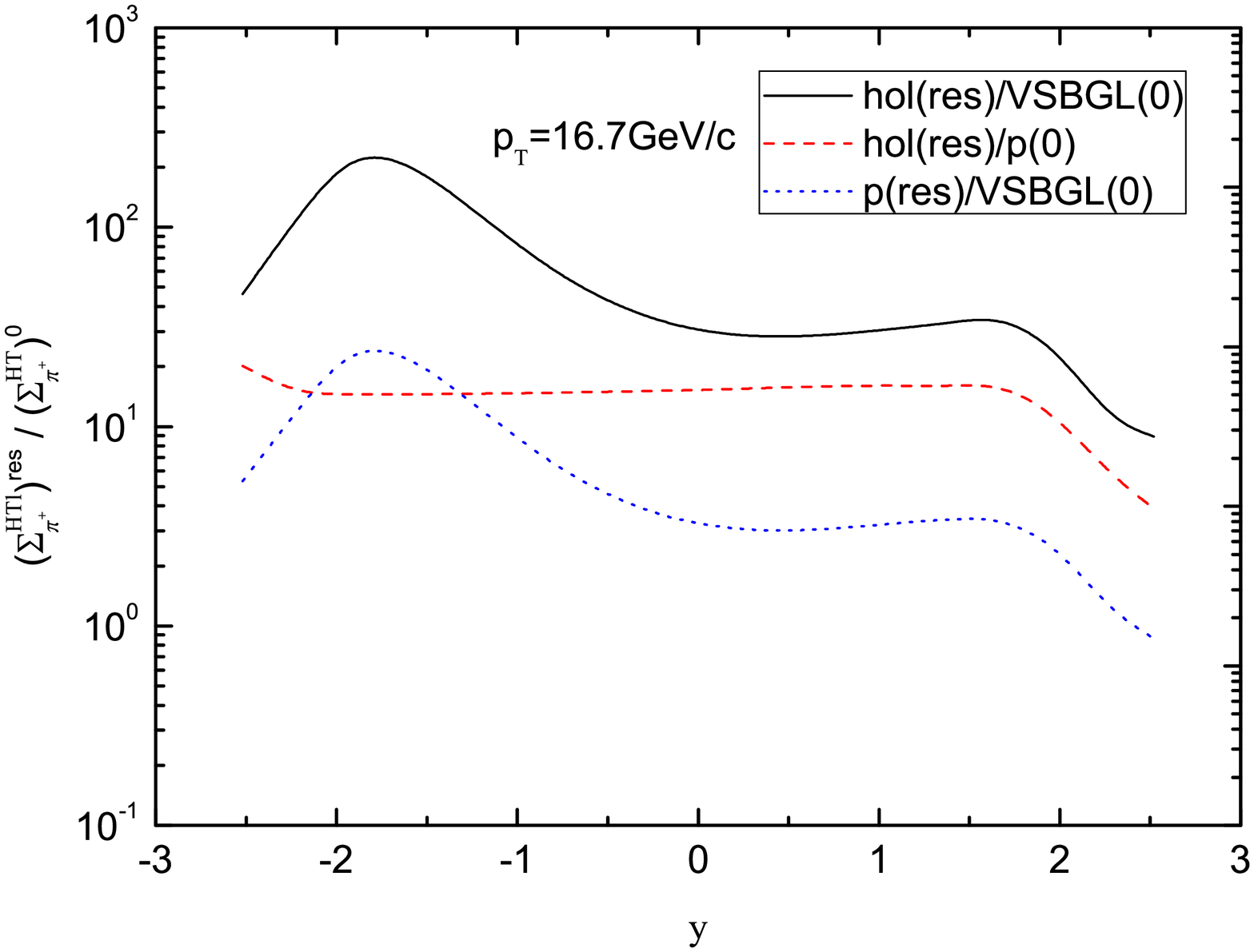}}
\vskip-0.2cm \caption{Ratio
$(\Sigma_{\pi^{+}}^{HT})^{res}/(\Sigma_{\pi^{+}}^{HT})^{0}$, as a
function of the $y$ rapidity of the pion at the  transverse momentum
of the pion $p_T=16.7\,\, GeV/c$, at the c.m. energy $\sqrt
s=209\,\, GeV$.} \label{Fig15}
\end{figure}

\end{document}